\documentstyle[preprint,aps,psfig]{revtex}

\def\lsim{\raise0.3ex\hbox{$\;<$\kern-0.75em\raise-1.1ex
\hbox{$\sim\;$}}}
\def\gsim{\raise0.3ex\hbox{$\;>$\kern-0.75em\raise-1.1ex
\hbox{$\sim\;$}}}
\begin{document}
\textheight = 23.0cm
\topmargin = -1.8cm
\tightenlines

\preprint{\vbox{\hbox{TMUP-HEL-0003}
\hbox{hep-ph/0004114}
\hbox{April 2000, Revised August 2000}
}}
\title{
Measuring Leptonic CP Violation by Low Energy Neutrino 
Oscillation Experiments 
}

\author{ 
Hisakazu Minakata~\thanks{E-mail: minakata@phys.metro-u.ac.jp}
}

\address{\sl
Department of Physics, Tokyo Metropolitan University \\
1-1 Minami-Osawa, Hachioji, Tokyo 192-0397, Japan, and \\
Research Center for Cosmic Neutrinos, 
Institute for Cosmic Ray Research, \\ 
University of Tokyo, Kashiwa, Chiba 277-8582, Japan}

\author{ 
Hiroshi Nunokawa~\thanks{E-mail: nunokawa@ifi.unicamp.br}
}

\address{\sl
Instituto de F\'{\i}sica Gleb Wataghin\\
Universidade Estadual de Campinas - UNICAMP\\
P.O. Box 6165, 13083-970 Campinas SP Brazil} 

\maketitle
\vspace{-0.4cm}
\hfuzz=25pt
\begin{abstract}
We uncover an interesting phenomenon that neutrino flavor transformation 
in slowly varying matter density imitates almost exactly that of vacuum 
neutrino oscillation under suitably chosen experimental parameters. 
It allows us to have relatively large CP violating measure 
$\Delta P \equiv P(\nu_{\mu} \rightarrow \nu_e) - 
P(\bar{\nu}_{\mu} \rightarrow \bar{\nu}_e)$ 
which is essentially free from matter effect contamination. 
We utilize this phenomenon to design a low-energy long-baseline 
neutrino oscillation experiment to measure the leptonic CP violating 
phase.

\end{abstract}
\pacs{PACS numbers:14.60.Pq,25.30.Pt}

\newpage
Exciting discovery of neutrino oscillation in atmospheric neutrino
observation \cite {SKatm} and the persistent discrepancy between 
the observed and the calculated flux of solar neutrinos 
\cite {solar} provide 
the strongest evidence for neutrino masses and lepton flavor mixing. 
Determination of all the mixing parameters, in particular the  
CP violating Kobayashi-Maskawa phase \cite{KM}, is one of the most 
challenging goals in particle physics. 

This is the third in a series of works \cite {MN97,MN98}
in which we intend to explore 
possible (and hopefully experimentally feasible) ways of measuring 
leptonic CP violation in neutrino oscillation experiments. 
For early references and recent works on CP violation (or 
equivalently T violation), see for example, Refs. \cite {early,KPTV} 
and Refs. \cite {AKS97,recent}, respectively. 
We focus in this paper the neutrino mass (difference) hierarchy suggested 
by the atmospheric neutrino observation and the MSW solutions 
\cite {MSW} of the solar neutrino problem \cite{AKS97}. 
In the standard three-flavor mixing scheme of neutrinos they exhaust
all the independent mass difference squared: 
from atmospheric neutrino data,  
$\Delta m_{13}^2 \approx \Delta m_{23}^2 = 
\Delta m_\textrm{atm}^2 \simeq (2-5)\times 10^{-3}\ \textrm{eV}^2$ 
and from solar neutrino data,  
$\Delta m_{12}^2 = \Delta m_\textrm{solar}^2 \simeq 
(4 - 10)\times10^{-6}\ \textrm{eV}^2\ \textrm{(SMA)},\ 
(2 - 20)\times 10^{-5}\ \textrm{eV}^2\ \textrm{(LMA)}$ or 
$(6 - 20)\times 10^{-8}\ \textrm{eV}^2\ \textrm{(LOW)}$
where SMA, LMA, and LOW denote the small mixing angle, 
large mixing angle and the low $\Delta m^2$ MSW solutions, 
respectively \cite{MSWfit,FLMP}. 
We use the notation in this paper as 
$\Delta m^2_{ij} \equiv m_j^2 - m_i^2$.

Many authors including us examined the question of how to separate 
the genuine CP violating effect due to the leptonic Kobayashi-Maskawa 
phase from the fake one induced by matter effect 
\cite{MN97,MN98,AKS97,recent}. 
In this paper we take a simple alternative strategy to look for 
a region of parameters in which the matter effect is "ignorable" 
in the first approximation. 
More precisely speaking, we will look for the solution of the 
question; is there region of tunable parameters in experiments, such as 
energy of neutrino beam, baseline length, etc. in which the neutrino 
oscillation probability, including its CP-odd term, are dominated 
by vacuum mixing effects?
(See below on what we mean precisely by "dominated by vacuum mixing 
effects".) 
We will answer the question in the positive and  
finally end up with a proposal of experiment which utilizes 
low-energy neutrino beam of $E \sim 100$ MeV and a megaton water 
Cherenkov detector to measure leptonic CP violation. 

We define the flavor mixing matrix $U$ as
$\nu_{\alpha}=U_{\alpha i}\nu_i$, where 
$\nu_{\alpha}(\alpha=e,\mu, \tau)$ and $\nu_i(i=1,2,3)$
stand for the gauge and the mass eigenstates, respectively.
We take for convenience the representation of $U$ as 
\begin{equation}
U = e^{i\lambda_7 \theta_{23}} \Gamma_{\delta}e^{i \lambda_5\theta_{13}}
e^{i\lambda_2\theta_{12}} 
\end{equation}
\begin{eqnarray}
=\left[
\matrix {c_{12}c_{13} & s_{12}c_{13} &  s_{13}\nonumber\\
-s_{12}c_{23}-c_{12}s_{23}s_{13}e^{i\delta} & 
c_{12}c_{23}-s_{12}s_{23}s_{13}e^{i\delta} & 
s_{23}c_{13}e^{i\delta} \nonumber\\
s_{12}s_{23}-c_{12}c_{23}s_{13}e^{i\delta} & 
-c_{12}s_{23}-s_{12}c_{23}s_{13}e^{i\delta} & 
c_{23}c_{13}e^{i\delta}\nonumber\\}
\right],
\label{eqn:CKM}
\end{eqnarray}
where $\lambda_i$ are SU(3) Gell-Mann's matrix and 
$\Gamma_{\delta} = diag (1, 1, e^{i\delta})$.

Let us first assume that the matter effect plays a minor role 
and consider the CP violating effect in vacuum.  
We will justify this assumption later.   
Under the mass difference hierarchy $\Delta m_{13}^2 \gg \Delta m_{12}^2$ 
which is implied by the solar and atmospheric neutrino data, 
the neutrino oscillation probability in vacuum can be written as
\begin{eqnarray}
P(\nu_\beta \rightarrow \nu_\alpha) 
&=& 
4 |U_{\alpha 3}|^2|U_{\beta 3}|^2
\sin^2 \left(\frac{\Delta_{13}L}{2}\right)
 -4 \mbox{Re} [U_{\alpha 1}U_{\alpha 2}^*U_{\beta 1}^*U_{\beta 2}]
\sin^2 \left(\frac{\Delta_{12}L}{2}\right)\nonumber\\ 
&& - 2J \sin (\Delta_{12}L) 
[1 - \cos(\Delta_{13}L)] 
+ 4J \sin (\Delta_{13}L) \sin^2 (\Delta_{12}L), 
\label{probvac}
\end{eqnarray}
where 
$\Delta_{ij} \equiv  \frac{\Delta m^2_{ij}}{2E} $. 
One of the most significant feature of (\ref{probvac}) is 
that CP violation comes in through $J$ defined by 
\begin{equation}
J_{\alpha\beta; i,j} \equiv \mbox{Im}
[U_{\alpha i}U_{\alpha j}^*U_{\beta i}^*U_{\beta j}]
\end{equation}
as it is the unique 
(in three-flavor mixing scheme) measure for CP violation 
as first observed by Jarlskog \cite{Jarlskog} in the case of quark mixing. 
It takes the form in the parametrization we introduced above as
$J = \pm c_{12}s_{12}c_{23}s_{23}c_{13}^2s_{13} \sin{\delta}$, 
where the sign is positive for $(e, \mu)$ and (1, 2) and $+(-)$ 
corresponds to their (anti-) cyclic permutations of $(\alpha, \beta)$ 
and $(i, j)$.

We first observe that from the expression of $J$, if any one of 
the mixing angles is extremely small or very close to $\pi/2$ 
there is little hope in detecting the leptonic CP violation.  
For this reason, we will not deal with the case of 
SMA MSW solar neutrino solution. 
See, however, a remark at the end of this paper.

We also notice that in order to have appreciable effect of 
CP violation, 
\begin{equation}
\Delta_{12} L = 0.26
\left(\frac{\Delta m_{12}^2}{10^{-5}\ \mathrm{eV}^2}\right)
\left(\frac{L}{1000\ \mathrm{km}}\right)
\left(\frac{E}{100\ \mathrm{MeV}}\right)^{-1}, 
\label{deltaL}
\end{equation}
should not be too small. 
If $\Delta m_{12}^2 \equiv \Delta m^2_\textrm{solar}$ is much smaller 
than $10^{-5}\ \mbox{eV}^2$ we have to either use a neutrino energy much 
lower than $\sim$ 100 MeV which would be impractical in 
accelerator experiments, or take baseline length much longer than 
1,000 km at the cost of lowering the beam flux. 
For this reason, our discussion can accommodate neither the just-so 
vacuum oscillation solution nor the LOW MSW solution.
Thus we assume in the following discussion that the nature chooses 
the LMA MSW solution to the solar neutrino problem.

If we restrict ourselves into relatively short baseline, i.e., 
less than 1,000 km, one can safely assume that linear approximation 
$\sin{(\Delta_{12}L)} \simeq \Delta_{12}L$ is valid. 
Therefore, in order that the CP 
violating (third) term in (\ref{probvac}) is relatively larger 
than the 2nd term we must focus on the region where 
$\Delta_{12}L$ is smaller but not too smaller than unity. 
This consideration naturally leads us to the option of low-energy 
neutrino oscillation experiments with energy of order $\sim$ 100 MeV.

Next we must justify our assumption that the matter effect 
plays a minor role even for the CP violating effect.
What we will show below is actually that whereas 
the Jarlskog factor and the energy eigenvalues are strongly modified 
by the matter effect it almost cancels out and does not show up 
in the observable quantities, the oscillation probabilities.  

We now discuss full system of three-flavor neutrino propagation 
in earth matter to understand the phenomenon of matter neutrino 
oscillation imitating vacuum oscillation. 
Toward this goal we develop an 
analytic framework based on perturbation theory under the adiabatic 
approximation \cite {KS99,Yasuda99}.
We start from the neutrino evolution equation in the flavor 
basis:
\begin{equation}
i\frac{d}{dx} 
\left[
\begin{array}{c}
\nu_e \\ \nu_\mu \\ \nu_\tau
\end{array}
\right] 
=
\left\{U \left[
\begin{array}{ccc}
0 & 0 & 0 \\
0 & \Delta_{12} & 0 \\
0 & 0 & \Delta_{13} 
\end{array}
\right] U^{\dagger}
+
\left[
\begin{array}{ccc}
a(x) & 0 & 0 \\
0 & 0 & 0 \\
0 & 0 & 0
\end{array}
\right]\right\}
\times
\left[
\begin{array}{c}
\nu_e \\ \nu_\mu \\ \nu_\tau
\end{array}
\right]
\label{evolution1}
\end{equation}
where $a(x)= \sqrt{2} G_F N_e(x)$ indicates the index of refraction 
with $G_F$ and $N_e$ 
being the Fermi constant and the electron number density, respectively.

We first note that at neutrino energy of $\sim$ 100 MeV there is a 
hierarchy among the relevant energy scales;
\begin{equation}
\Delta_{13} \equiv \frac{\Delta m_{13}^2}{2E} \gg a \sim
\frac{\Delta m_{12}^2}{2E} \equiv \Delta_{12}
\label{hierarchy}
\end{equation}
The latter relationship holds because
\begin{equation}
\frac{a(x)}{\Delta m^2/2E} = 2.1 
\left(\frac{\rho}{2.7\ \mathrm{g/cm}^3}\right)
\left(\frac{Y_e}{0.5}\right)
\left(\frac{\Delta m^2}{10^{-5}\ \mathrm{eV}^2}\right)^{-1}
\left(\frac{E}{100\ \mathrm{MeV}}\right).
\end{equation}
Thanks to the mass hierarchy (\ref{hierarchy}) we can formulate 
the perturbation theory. 

We rewrite the evolution equation (\ref{evolution1}) as \cite{KP87} 
\begin{equation}
i \frac{\mathrm{d}}{{\mathrm{d}}x} \tilde{\nu_{\alpha}} = 
(H_0 + H')_{\alpha \beta}, 
\tilde{\nu_{\beta}}
\end{equation}
where $\tilde{\nu}$ is defined by
\begin{equation}
\tilde{\nu_{\alpha}} =
\left[
e^{-i\lambda_5\theta_{13}} \Gamma_{\delta}^{+} e^{-i\lambda_7\theta_{23}}
\right]_{\alpha\beta} \nu_{\beta}, 
\end{equation}
and 
$H_0 = diag (0, 0, \Delta_{13})$ and 
\begin{equation}
H' = \Delta_{12} \left[
\begin{array}{ccc}
s_{12}^2 & c_{12}s_{12} & 0 \\
c_{12}s_{12} & c_{12}^2 & 0 \\
0 & 0 & 0 
\end{array}
\right] 
+ a(x) \left[
\begin{array}{ccc}
c_{13}^2 & 0 & c_{13}s_{13} \\
0 & 0 & 0 \\
c_{13}s_{13} & 0 & s_{13}^2
\end{array}
\right]. 
\end{equation}

The diagonalization of the 2$\times$2 submatrix gives rise to the 
energy eigenvalues $h_i$ and the matter enhanced $\theta_{12}$ as 
\begin{equation}
h_{1,2} = \frac{1}{2}
\left[
c_{13}^2 a(x) + \Delta_{12} \mp 
\sqrt{(\cos 2\theta_{12}\Delta_{12} - ac_{13}^2)^2 + 
\Delta_{12}^2 \sin^2 2\theta_{12}}
\right],
\label{energyM}
\end{equation}
\begin{equation}
\sin 2\theta_{12}^M =
\displaystyle\frac{\sin 2\theta_{12}}{\sqrt{(\cos 2\theta_{12} - 
\frac{a}{\Delta_{12}}c_{13}^2)^2 + \sin^2 2\theta_{12}}}
\label{angleM}
\end{equation}
The matter ``suppressed'' $\bar{\theta}_{12}^M$ for antineutrinos 
and the energy eigenvalues can be defined analogously by flipping 
the sign of $a$ in (\ref{energyM}) and (\ref{angleM}). 
The resonance condition is satisfied in neutrino channel \footnote
{
We note that one can take $\Delta m^2_{12}$ always positive in the 
MSW mechanism as far as $\theta_{12}$ is taken in its full range
[0, $\pi/2$]\cite {FLMP}.
} 
at 
\begin{equation}
c_{13}^2 a(x) = \cos 2\theta_{12}\Delta_{12}
\end{equation}
which leads to the resonance energy of order of 100 MeV, 
\begin{equation}
E_r = 48.6 \cos 2\theta_{12}
\left(\frac{\Delta m_{12}^2}{10^{-5}\mathrm{eV}^2}\right)
\left(\frac{\rho}{2.7\mathrm{g/cm}^3}\right)^{-1}
\left(\frac{Y_e}{0.5}\right) \mathrm{MeV}
\label{resenergy}
\end{equation}
for small value of the remaining vacuum mixing angle, 
$\theta_{13} \ll 1$. 
For the best fit values of the parameter of the LMA MSW solution, 
$\Delta m^2_{12} \sim (3.0-4.0) \times 10^{-5}$ eV$^2$ and 
$\sin^2 2\theta_{12} \sim 0.8$ \cite{MSWfit}, 
we obtain $E_r \sim 60-80$ MeV.

The key to the matter enhanced $\theta_{12}$ mechanism is the degeneracy 
of the zeroth order "energy" eigenvalue in the first 2$\times$2 subspace 
of $H_0$. The degenerate perturbation theory dictates that one has to 
first diagonalize $H_0 +H'$ in this subspace to obtain the first-order 
corrected energy eigenvalues and the {\it zeroth-order} wave function. 
This last point is crucial. It gives rise to the matter enhanced CP 
violating effect that is free from the suppression by energy denominator. 
To our knowledge, this is the unique case of having CP violating effect 
which is not suppressed by any hierarchical ratios such as 
$\Delta m_{12}^2/\Delta m_{13}^2$

While our formalism is exactly the same as the one developed in  
Ref. \cite {KS99} the interpretation of the physical 
phenomena that occur is quite different from theirs. 
Also the nature of the resonance is quite different in 
large mixing angles. 
The resonance width can be estimated as 
$\frac {\delta E}{E_r} = \tan 2\theta_{12}$
which means that $\delta E \simeq 2 E_r \simeq$ 120 MeV.  
Therefore, the resonances are so broad that they lose the identity 
as sharp resonances. 

We now show that despite the fact that $\theta_{12}$ could be
strongly modified in matter, especially for small vacuum $\theta_{12}$, 
the system mimics the vacuum neutrino oscillation
even at the resonance where the effect of matter could be 
maximal. 
To understand this point we calculate the neutrino and antineutrino 
conversion probabilities $P(\nu_{\mu} \rightarrow \nu_e)$ and 
$P(\bar{\nu}_{\mu} \rightarrow \bar{\nu}_e)$. 
Since the matter enhanced mixing angle $\theta_{12}^M$ just replaces 
$\theta_{12}$ in zeroth-order wave function it is straightforward to 
compute neutrino oscillation probabilities under the adiabatic 
approximation. 
They are nothing but the oscillation probabilities in vacuum but with 
$\theta_{12}$ replaced by $\theta_{12}^M$, and $\Delta_{ij}$ by
integrals over the energy eigenvalues, $h_{1,2}$ and 
$h_3 \simeq \Delta_{13}$.
For example, the appearance probability 
$P(\nu_{\mu} \rightarrow \nu_e)$ 
reads 
\begin{eqnarray}
P(\nu_{\mu} \rightarrow \nu_e) &=& 
4 s_{23}^2 c_{13}^2 s_{13}^2 \sin^2 (\frac{1}{2}\Delta_{13}L) \nonumber\\
&& + 
c_{13}^2\sin 2\theta_{12}^M \left[ 
(c_{23}^2 - s_{23}^2 s_{13}^2) \sin 2\theta_{12}^M 
+ 2c_{23}s_{23}s_{13}\cos{\delta}\cos{2\theta_{12}^M} \right]\nonumber\\
&& \hskip 2 cm \times
\sin^2
\left[
{\frac{1}{2}\sqrt{(\cos 2\theta_{12} - 
\frac{a}{\Delta_{12}}c_{13}^2)^2 + \sin^2 2\theta_{12}}\Delta_{12}L}
\right]\nonumber\\
&& - 
2J_M (\theta_{12}^M, \delta)
\sin
\left[
{\sqrt{(\cos 2\theta_{12} - 
\frac{a}{\Delta_{12}}c_{13}^2)^2 + \sin^2 2\theta_{12}}\Delta_{12}L}
\right]
\label{probnu}
\end{eqnarray}
where $J_M$ is the matter enhanced Jarlskog factor, 
$J_M (\theta_{12}^M, \delta) 
=\cos \theta_{12}^M \sin \theta_{12}^M c_{23}s_{23}c_{13}^2
s_{13}\sin \delta$, and we have averaged the rapidly oscillating 
piece driven by $\Delta_{13}$ in the CP violating term.
The antineutrino transition probability 
$P(\bar{\nu}_{\mu} \rightarrow \bar{\nu}_e)$
is given by the same expressions as above but replacing 
$\theta_{12}^M$ by $\bar{\theta}_{12}^M$ and $\delta$ by 
$- \delta$. 

Notice that at relatively short baseline, $L <$ 1,000 km or so, 
the approximation $\sin x \simeq x$ 
is valid. Then, 
the expressions of the oscillation probabilities approximately 
reduce to those in the vacuum because 
\begin{equation}
\sin 2\theta_{12}^M (\mbox{or}, \bar{\theta}_{12}^M)
\sqrt{(\cos 2\theta_{12} \mp \frac{a}{\Delta_{12}}c_{13}^2)^2 + 
\sin^2 2\theta_{12}}\Delta_{12} = 
\sin 2\theta_{12} \Delta_{12}. 
\end{equation}
Only mild $a$-dependence would remain due to the 
$\cos \delta$ term in (\ref{probnu}). 
Hence, as long as $\Delta_{12} L $ is small, 
the oscillation probabilities of neutrinos and antineutrinos
in matter imitate those in vacuum, independent of the mixing parameters 
and neutrino energy. 

We verify by numerical computations without using analytic expression in 
eq. (\ref{probnu}) 
that the matter effect in fact cancels out.
We tentatively take the values of the parameters as 
$\Delta m^2_{13} = 3 \times 10^{-3}$ eV$^2$, 
$\sin^22\theta_{23} = 1.0$, 
$\Delta m^2_{12} = 2.7 \times 10^{-5}$ eV$^2$, 
$\sin^22\theta_{12} = 0.79$,
$\sin^22\theta_{13} = 0.1$ and 
the Kobayashi-Maskawa phase $\delta = \pi/2$. 
Our choice of $\theta_{13}$ is within the newer 
CHOOZ bound \cite{CHOOZ}, but it may still be an optimistic one.
With these parameters the Jarlskog factor in vacuum is given by 
$J = 0.035$, a small but non-negligible value.
If we take $\delta = - \pi/2$ it corresponds, in a good approximation, 
to interchanging $\nu$ and $\bar{\nu}$ because the matter effect is 
only minor.

\begin{figure}[ht]
\vglue 1.8cm 
\hglue -1.0cm 
\centerline{\protect\hbox{
\psfig{file=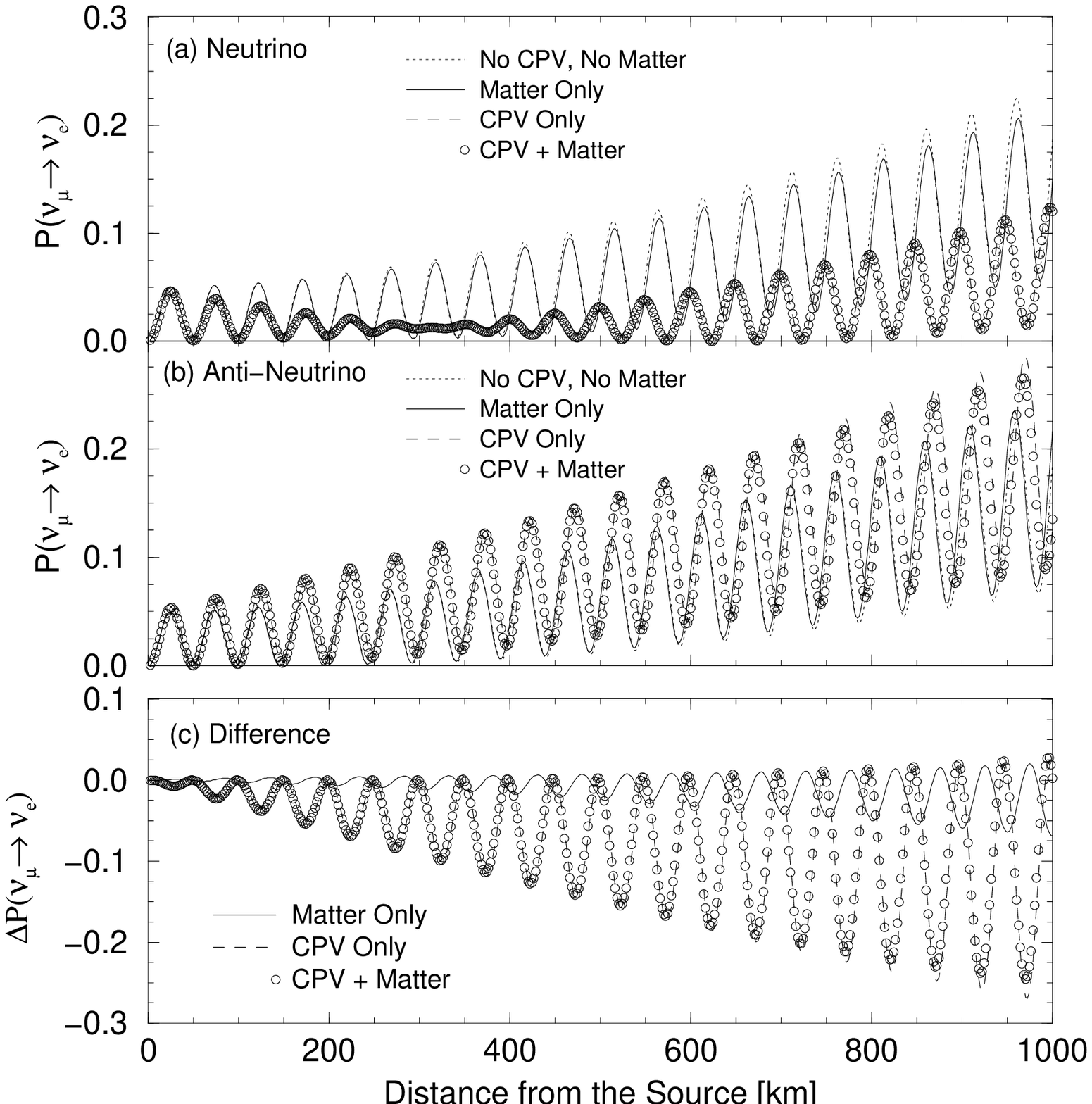,height=16cm,width=18.cm}
}}
\vglue 1.4cm 
\caption{
Oscillation probability for (a) neutrinos, $P(\nu_\mu\to\nu_e)$,  
(b) anti-neutrinos, $P(\bar{\nu}_\mu\to\bar{\nu}_e)$,  
and (c) their difference,
$\Delta P(\nu_\mu\to\nu_e) \equiv 
P(\nu_\mu \to \nu_e) - P(\bar{\nu}_\mu\to\bar{\nu}_e)$ 
with fixed neutrino energy $E_\nu = $ 60 MeV, 
are plotted as a function of distance from the source. 
The mixing parameters are fixed to be 
$\Delta m^2_{13} = 3 \times 10^{-3}$ eV$^2$, 
$\sin^22\theta_{23} = 1.0$, 
$\Delta m^2_{12} = 2.7 \times 10^{-5}$ eV$^2$, 
$\sin^22\theta_{12} = 0.79$,
$\sin^22\theta_{13} = 0.1$ and 
$\delta = \pi/2$. 
We take the matter density as $\rho = 2.72$ g/cm$^3$ and 
the electron fraction as $Y_e$ = 0.5. 
}
\label{Fig1}
\end{figure}

In Fig. \ref{Fig1} we plot the oscillation probabilities for neutrino 
and anti-neutrino, and their difference, 
$\Delta P \equiv P(\nu_{\mu} \rightarrow \nu_e) - 
P(\bar{\nu_{\mu}} \rightarrow \bar{\nu_e})$, 
as a function of 
distance from the source with the neutrino energy $E = 60$ MeV, 
for both 
$\nu_{\mu} \rightarrow \nu_e$ and 
$\bar{\nu_{\mu}} \rightarrow \bar{\nu_e}$.
It corresponds to the resonance energy in the neutrino channel.
Insensitivity of the transition probability to the matter effect and 
how well it mimics the vacuum oscillation probability is clearly 
displayed in this figure. 
Although we sit on at the resonance energy of neutrino channel 
the features in the resonant neutrino flavor conversion cannot be 
traced in Fig. \ref{Fig1}, but rather we observe the one very much similar 
to the vacuum oscillation. 
We have also checked that the matter effect is small
even if $\cos \delta$ term in (\ref{probnu}) is non-zero. 
The feature that it is a product of two harmonics with quite 
different frequencies can be understood by (\ref{probvac}).
We note that a large amplitude of $|\Delta P|$ 
(as large as $\sim$ 0.3) in Fig. \ref{Fig1}
can be qualitatively understood from the third term in 
our approximated analytic expression in eq. (\ref{probvac}) 
which implies that $|\Delta P|$ can be as large as 
$8J (\sim 0.28)$ at $\Delta_{13}L = \pi$.

Now we discuss the possible experiments which utilize 
the imitating vacuum mechanism to measure leptonic CP violation.
Measurement of CP violation at a few $\%$ level at neutrino 
oscillation experiments at $E \simeq 100$ MeV leaves 
practically the unique channel $\nu_{\mu} \rightarrow \nu_e$.
Experimentally, what we can actually measure is the number 
of events not the oscillation probability itself. 
Here we consider the ratio of the expected number
of events due to $\nu_{\mu} \rightarrow \nu_e$ 
and $\bar{\nu_{\mu}} \rightarrow \bar{\nu_e}$ 
reactions in order to quantify the effect of CP violation. 

In such low energy appearance experiment 
we must circumvent the following difficulties; 
(1) smaller cross sections (2) lower flux due to larger beam opening 
angle, $\Delta \theta \simeq 1$ (E/100 MeV)$^{-1}$ radian. 
Therefore, we are invited to the idea of the baseline as short as 
possible, because the luminosity decreases as $L^{-2}$ as baseline 
length grows. But of course, a baseline as long as possible is 
preferable to make the CP violating effect [the last term in 
(\ref{probnu})] maximal. Thus we have to compromise.
It appears to us that for $E \sim 100 $ MeV, 
the baseline of $\sim$ 30-50 km seems preferable, 
as we will see below. 

Detection of low-energy neutrinos at a few $\%$ level accuracy 
requires supermassive detectors. Probably the best 
thinkable detection apparatus is the water Cherenkov detector of 
Superkamiokande type. Let us estimate the expected number of events 
at a megaton detector placed at $L=100$ km. We assume that the neutrino 
beam flux 10 times as intense as (despite the difference in energy) 
that of the design luminosity in K2K experiment \cite {Nishikawa}. 
In the future it is expected that an 100 times more intense 
proton flux than KEK-ps seems possible at Japan Hadron Facility 
\cite {Mori} but we assume only 10 times larger flux considering 
the fact that lower energy beam is more difficult to prepare.

The dominant $\nu_e$-induced reaction in water at around $E=100$ MeV 
is not the familiar $\nu_e-e$ elastic scattering but the reaction on 
$^{16}O$,  $\nu_e ^{16}O \rightarrow e^{-} F$ \cite {Haxton}.
The cross section of the former reaction is about 
$\sigma(\nu_e e \rightarrow \nu_e e) =
0.93 \times 10^{-42}
\displaystyle\left(\frac{E}{100\ \mathrm{MeV}}\right) 
\mathrm{cm}^2$, while the latter 
is 
$\sigma(\nu_e ^{16}O \rightarrow e^{-} F) \simeq 10^{-39} 
\mathrm{cm}^2$ at $E=100$ MeV \cite {Haxton}, 
which is a factor of 1000 times 
larger.\footnote{
Notice that the energy dependence of $\nu_e ^{16}O$ reaction  
is very steep below $E=100$ MeV. For example, it is smaller by 
an order of magnitude at $E=60$ MeV. Because the energy dependence 
of $\Delta P$ is mild we are invited to take larger energy than the 
resonance energy.}
But the number of oxygen in water is 1/10 of the number of electrons. 
So in net the number of events due to the reaction 
$\nu_e ^{16} O \rightarrow e^{-} F$ is larger than that of 
$\nu_e e$ elastic scattering by a factor of 100. The number of 
$^{16}O$ in detector is given by $3.34 \times 10^{31}$
per kton of water. 
The neutrino flux at the detector located at $L=250$ km is, 
by our assumption, 10 times more intense 
than the neutrino flux at Superkamiokande in K2K experiment.
The latter is, roughly speaking, 
$3 \times 10^{6} 
\displaystyle\left(\frac{\mathrm{POT}}{10^{20}}\right)$ 
cm$^{-2}$ where 
POT stands for proton on target.
Therefore, the expected number of events $N$ assuming 100 $\%$ 
conversion of $\nu_{\mu}$ to $\nu_e$ is given by 
\begin{equation}
N \simeq 6300
\ \displaystyle\left(\frac{L}{100\ \mathrm{km}}\right)^{-2}
\displaystyle\left(\frac{V}{1\ \mathrm{Mton}}\right) 
\displaystyle\left(\frac{\mathrm{POT}}{10^{21}}\right). 
\label{eventNo}
\end{equation}

In the antineutrino channel, the dominant reaction is 
$\bar{\nu}_e p \rightarrow e^+ n$ with cross section
$\sigma \simeq 0.4 \times 10^{-39} \mathrm{cm}^2$ at $E=100$ MeV. 
The event number due to this reaction, assuming the same 
flux of $\bar{\nu}_{\mu}$ as $\nu_{\mu}$, is similar to that of 
(\ref{eventNo}) because the cross section is about half but 
there are two free protons per one oxygen.  
There is additional oxygen reaction 
$\bar{\nu}_e ^{16}O \rightarrow e^{+} N$ 
with approximately factor 3 smaller cross section than that 
of $\nu_e ^{16}O$ \cite {Haxton}.

\begin{figure}[ht]
\hglue -1.0cm 
\centerline{\protect\hbox{
\psfig{file=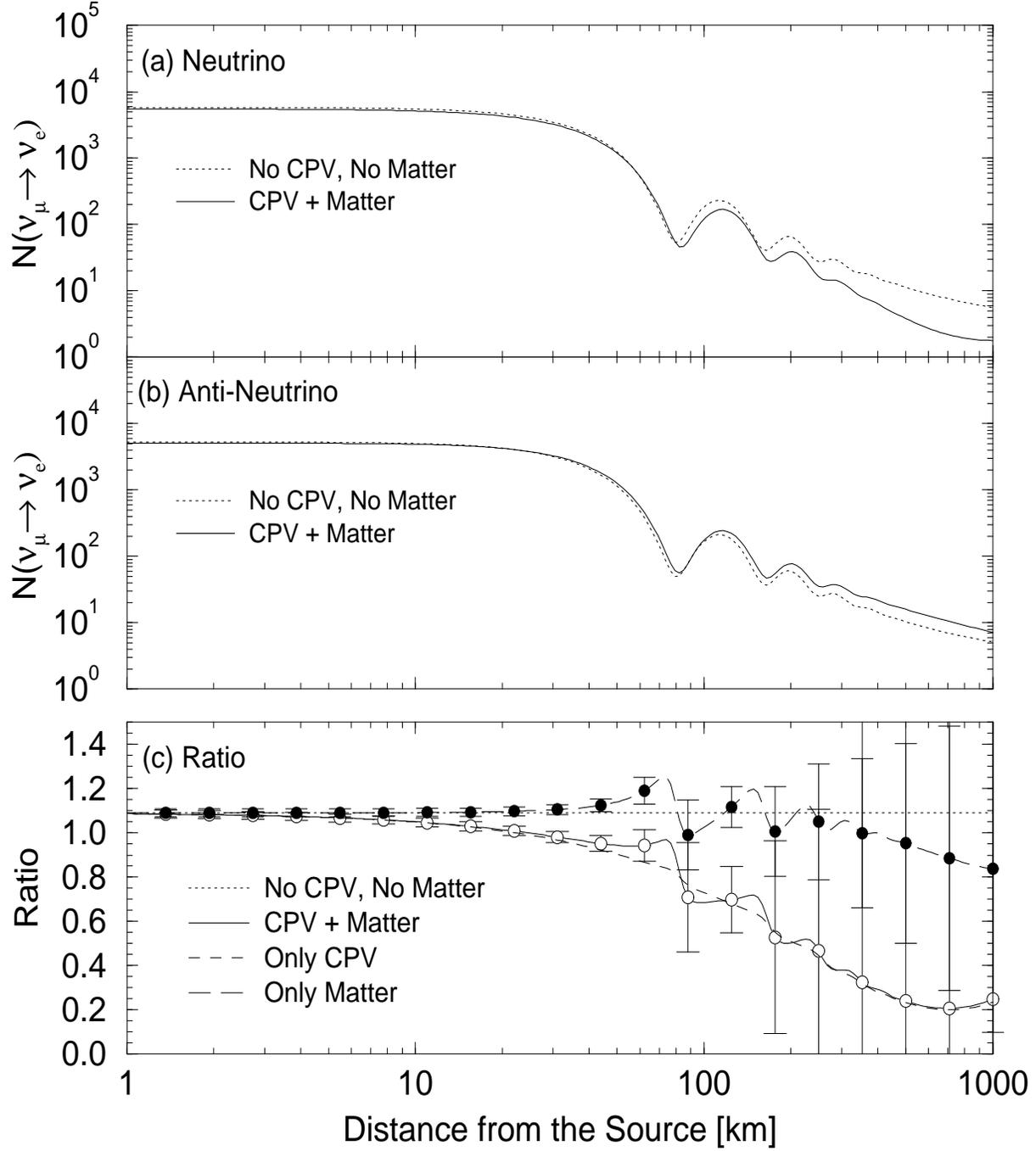,height=16cm,width=18.cm}
}}
\vglue 1.0cm 
\caption{
Expected number of events for 
(a) neutrinos, $N(\nu_\mu\to\nu_e)$,  
(b) anti-neutrinos, $N(\bar{\nu}_\mu\to\bar{\nu}_e)$,  
and (c) their ratio 
$R\equiv N(\nu_\mu \to \nu_e)/N(\bar{\nu}_\mu\to\bar{\nu}_e)$ 
with a Gaussian type neutrino energy beam with 
$\langle E_\nu \rangle = $ 100 MeV with $\sigma$ = 10 MeV 
are plotted as a function of distance from the source. 
Neutrion fluxes are assumed to vary as $\sim 1/L^2$ 
in all the distance range we consider. 
The mixing parameters as well as the electron number density 
are fixed to be the same as in Fig. 1. 
The error bars are only statistical.
}
\label{Fig2}
\end{figure}

In order to estimate the optimal distance, we compute the expected 
number of events in neutrino and anti-neutrino channels as well as 
their ratios as a function of distance with taking into account of 
neutrino beam energy spread. 
For definiteness, we assume that the average energy of neutrino 
beam $\langle E \rangle$ = 100 MeV and beam energy spread of 
Gaussian type with width $\sigma_E = 10$ MeV. 
We present our results in Fig. \ref{Fig2}.

If there is no matter and CP violating phase effects the ratio 
$R \equiv N(\nu_\mu \to \nu_e)/ N(\bar{\nu}_\mu \to \bar{\nu}_e)$ 
must not vary with the distance but must be a constant 
which is $ \phi_\nu\langle \sigma_\nu \rangle/ 
\phi_{\bar{\nu}} \langle \sigma_{\bar{\nu}} \rangle$
where $\phi_\nu$ ($\phi_{\bar{\nu}}$) and 
$\sigma_\nu$ ($\sigma_{\bar{\nu}}$) are total flux and
cross section averaged over neutrino (anti-neutrino) energy, 
respectively.  
Taking the error (only statistical one for simplicity) 
into account, $L$ must not be too short or too long.
The optimal distance turned out to be rather short, $L \sim 30-50 $ km, 
much shorter than that of the current long-baseline experiments.
(see Fig. 2.)
At such distance the significance of the CP violating
signal is as large as 3-4 $\sigma$.

Roughly speaking, the deviation of $R$ from the case without 
CP violating effect, $\Delta R$, is 
proportional to $\Delta P$ but 
what is relevant is the quantity 
$\Delta P  \times N$ ($N$ is the expected number 
of events in the absence of CP violating effect)
which signals if the CP violating effect is 
significant or not.
From eq. (\ref{probvac}) we can see that when the distance 
is small so that $\Delta_{13}L \ll 1$, the terms proportinal 
to $J$ (or $\Delta P$) behaves like $\Delta P \propto (L/E)^3$. 
This implies that assuming $\phi_\nu \propto 1/L^2$, 
$\Delta P \times N \propto \Delta P\sigma(E)/L^2
\propto \sigma(E)L/E^3 $ is an increasing function of $L$. 
When the distance is larger and $\Delta_{13}L$ is not small,
$\Delta P \propto (L/E)$ and this implies that 
$\Delta P \times N \propto \sigma(E)E/L$ 
is a decreasing function of $L$.

Hence,  the significance of CP violating effect is expected to 
be maximal at some distance $L$ where $\Delta P \times N$ takes
the largest value. 
In our case it corresponds to $L \sim 30-50$ km for 
the mixing parameters and neutrino energy we consider, as 
indicated in Fig. 2.

Let us also make some comments on the validity of our choice of low 
neutrino energy (as low as $\sim 100$ MeV) from the view point 
of the expected number of events. 
Suppose that the relevant cross section $\sigma(E)$ behaves 
like $\sigma(E) \propto E^\alpha$ and we can neglect 
matter effect. 
In this case, $N$ is $\sim f(L/E) \times E^{(\alpha-2)}$ where
$f(L/E)$ is some function of $L/E$. 
From this we can conclude that if $\alpha = 2$ the same 
number of events can be expected as long as the value 
of $L/E$ is the same if the same 
intensity of neutrino beam is assumed
because the increase in the cross section with higher 
energy can be compensated by the decrese of the flux due to 
larger distance.  
On the other hand, if $\alpha \ne 2$ such scaling law does not
apply.  If $\alpha > 2$ ($\alpha < 2$) the one can expect 
larger number of events with higher (lower) energy 
if the value of $L/E$ is kept constant.  
For neutrino energy above 100 MeV but below 1 GeV, for neutrno 
channel, roughly speaking $\alpha \sim 2$ whereas for anti 
neutrino channel $\alpha \sim 1$ \cite{cross}. 
Therefore, our choice of lower energy seems to be preferable.  
For energy well above GeV, $\alpha \sim 1$ for both
channels and lower energy is preferable.  

We conclude that the experiment is quite feasible under such intense 
neutrino beam and a megaton detector. Fortunately, the possibility 
of constructing a megaton water Cherenkov detector is already 
discussed by the experimentalists \cite{Nakamura}.

Finally, a few remarks are in order:

\noindent
(1) Imitating vacuum mechanism does work also for the SMA solar 
MSW solution after averaging over the rapid oscillations due to 
larger $\Delta m^2$. Because of the tiny vacuum angle, however, the 
number of events is smaller by a factor of $\sim 2500$ than the 
LMA case and the experiment does not appear feasible with the 
same apparatus. 

\noindent
(2) An alternative way of measuring CP violation is the multiple 
detector method \cite{MN97} which may be inevitable if 
either one of $\nu$ or $\bar{\nu}$ beam is difficult to prepare. 
It utilizes the fact that the first, the second, and the third terms 
in the oscillation probability (\ref{probnu}) have different $L$ 
dependences, $\sim L$-independent 
(after averaging over energy spread of the neutrino beam), 
$\sim L^2$, and $\sim L$, respectively, 
in the linear approximation.\footnote
{
The minimal two detector methods \cite {MN97} is most powerful 
if the far (near) detector is placed at 
$\Delta_{12} = \frac{3\pi}{2} (\frac{\pi}{2})$ 
because then the difference 
of the CP violating term in (\ref{probnu}) between 
the far and the near detectors is maximal. However, it requires the 
baseline length of far detector as 
$L=6,300 
\displaystyle\left(\frac{\Delta m_{12}^2}
{3 \times 10^{-5}\mbox{eV}^2}\right)^{-1}
\left(\frac{E}{100\ \mbox{MeV}}\right)$ km 
which may or may not be too large to realize.
}

\noindent
(3) Intense neutrino beams from muon storage ring at low energies, 
proposed as PRISM, Phase-Rotation Intense Secondary Mesons \cite {Kuno}, 
would be an ideal source for neutrinos for the experiment proposed in 
this paper. Of course, it requires identification of 
$\bar{\nu_e}$ from $\nu_e$ by some methods, e.g., 
by adding Chlorine ($^{35}$Cl) into the Water Cherenkov detector 
to make it sensitive to the characteristic $\sim 8$ MeV $\gamma$ 
rays arising from the absorption of neutron into the Chlorine 
followed by $\bar{\nu}_e p \rightarrow e^+ n$ reaction. 

\noindent
(4) We would like to urge experimentalists to think more about the 
better supermassive detection apparatus than water Cherenkov for 
highly efficient and accurate measurement of low energy neutrinos.

\acknowledgments 

We thank Takaaki Kajita, Masayuki Nakahata and Kenzo Nakamura 
for informative discussions on detection of low energy neutrinos.
One of us (HM) thanks Marcelo M. Guzzo and the department of 
Cosmic Ray and Chronology at Gleb Wataghin Physics Institute 
in UNICAMP for their hospitality during his visit when the main 
part of this work was done. 
This work was supported by the Brazilian funding agency 
Funda\c{c}\~ao de Amparo \`a Pesquisa do Estado de S\~ao Paulo (FAPESP), 
and by the Grant-in-Aid for Scientific Research in Priority Areas No. 
11127213, Japanese Ministry of Education, Science, Sports and Culture.


\newpage
\vglue 1.0cm 

\end{document}